\documentclass{article}
\usepackage[utf8]{inputenc}
\usepackage{jheppub}
\usepackage{graphicx}
\usepackage{subfig}
\usepackage{float}
\usepackage{amsmath}
\usepackage{amssymb}
\usepackage{amsthm}
\usepackage{latexsym}
\usepackage{dcolumn}
\title{Impact of uncertainties in the halo velocity profile on direct detection of sub-GeV dark matter}

\author[a]{Andrzej Hryczuk,}
\author[b]{Ekaterina Karukes,}
\author[b,a]{Leszek Roszkowski,}
\author[b]{Matthew Talia}

\affiliation[a]{National Centre for Nuclear Research,\\
  Pasteura 7, 02-093 Warsaw, Poland}
\affiliation[b]{AstroCeNT, Nicolaus Copernicus Astronomical Center Polish Academy of Sciences,\\ul. Rektorska 4, 00-614 Warsaw, Poland}

\emailAdd{andrzej.hryczuk@ncbj.gov.pl}
\emailAdd{ekarukes@camk.edu.pl}
\emailAdd{leszek.roszkowski@ncbj.gov.pl}
\emailAdd{mtalia@camk.edu.pl}

\abstract{
We use the state-of-the-art high-resolution cosmological simulations by IllustrisTNG to derive the velocity distribution and local density of dark matter in galaxies like our Milky Way and find a substantial spread in both quantities. Next we use our findings to examine the sensitivity to the dark matter velocity profile of underground searches using electron scattering in germanium and silicon targets. We find that sub-GeV dark matter search is strongly affected by these uncertainties, unlike nuclear recoil searches for heavier dark matter, especially in multiple electron-hole modes, for which  the sensitivity to the scattering cross-section is also weaker. Therefore, by improving the sensitivity to lower ionization thresholds not only projected sensitivities will be boosted but also the dependence on the astrophysical uncertainties will become significantly reduced.
}

\begin{document}
\maketitle
\section{Introduction}
\label{sec:intro}

Our limited understanding of the dark matter (DM) halo structure, both local and Galactic, constitutes one of the largest sources of uncertainty in direct and indirect DM searches. As astrophysical observations present many challenges in determining the mass, density, shape and velocity profile of dark matter (see, e.g.,~\cite{Karukes:2019jxv,OHare:2019qxc}, where the latter is based on the recent results from the Gaia satellite), the method of choice for studying DM halo properties has often been based on numerical simulations of large scale structures. These, however, until recently suffered from lacking to account for the baryonic component, or an oversimplified implementation of its role. Only very recently progress in hydrodynamical simulations with state-of-the-art baryonic physics effects included allowed for sufficient precision to accurately simulate Milky Way (MW)-like galaxies.

In this paper we use the cosmological simulations of IllustrisTNG  \cite{Pillepich:2017jle,Nelson:2017cxy,Springel:2017tpz} to infer local properties of the DM halo structure of MW-like galaxies and subsequently apply our findings to examine the impact of the estimated astrophysical uncertainties on direct DM searches. Until recently significant effort was devoted to studying the implications of different assumptions about the local DM density and the velocity profile for the direct detection searches of heavy -- of a few GeV, or more -- DM based on nuclear recoils (see, e.g., a detailed discussion in \cite{Peter:2013aha}). Even though our results can be applied to this case as well, in this work we focus on a previously unexplored impact on direct detection of sub-GeV dark matter via electron scattering. As  we  shall  see, in this detection technique DM halo properties, especially its velocity distribution, start to play a significant role.

The search program for dark matter detection has widened up considerably in recent years. This is especially true on the direct detection front, where the primary focus had been on WIMPs (weakly interacting massive particles) with mass around and above the weak scale. More recently a lot of effort has been dedicated to developing novel detection techniques applicable to sub-GeV DM. These include, but are not limited to electron transitions: in atoms and semiconductors~\cite{Graham:2012su,Essig:2015cda,Hochberg:2016ntt,Derenzo:2016fse,Hochberg:2016sqx,Bloch:2016sjj,Essig:2017kqs,Kadribasic:2017obi,Heikinheimo:2019lwg}, superconductors~\cite{Hochberg:2015pha,Hochberg:2015fth,Hochberg:2016ajh}, topological insulators \cite{Marsh:2018dlj,Liang_2018} and Dirac materials~\cite{Hochberg:2017wce,Coskuner:2019odd,Geilhufe:2019ndy}. Sub-GeV DM can be searched for also via the Migdal effect~\cite{Ibe:2017yqa,Dolan:2017xbu,Baxter:2019pnz,Essig:2019xkx}, single phonon~\cite{Knapen:2017ekk,Griffin:2018bjn} and magnon~\cite{Trickle:2019ovy} excitations in crystals, in superfluid helium~\cite{Schutz:2016tid,Knapen:2016cue}, through dissociation or excitation in molecules~\cite{Essig:2016crl,Essig:2019kfe} and also through other methods~\cite{Kouvaris:2016afs,Cavoto:2017otc,Alonso:2018dxy,Blanco:2019lrf} proposed recently.

This paper consists of two main parts: in Sec.~\ref{sec:illustris} we construct and discuss several DM profiles inferred from the Illustris simulation, while in Sec.~\ref{sec:subGeV} we apply our findings to study their effect on electron recoil searches of DM with sub-GeV mass in semiconductor targets. Section \ref{sec:conclusions} provides discussion of the results and conclusions.

\section{Dark matter velocity distributions from IllustrisTNG simulation}
\label{sec:illustris}

As outlined in the introduction, dark matter velocity distribution is a crucial ingredient for estimating direct dark matter detection in the Milky Way. Most studies assume that the dark matter halo model follows the Standard Halo Model (SHM) \cite{Drukier:1986tm}, either for simplicity or to provide a platform for comparison among different experiments. The SHM is an isothermal, spherical dark halo with an isotropic Maxwell-Boltzmann (MB) dark matter velocity distribution with the dispersion velocity $v_0$ (taken to be the local circular speed, hereinafter referred to as \textit{peak speed}) in the Galactic rest frame and truncated at the escape speed from the Galaxy,
\begin{equation}
  f_{\rm gal}({\bf v}) = \left \{
  \begin{aligned}
    &N\,\exp\left({-{\bf v}^2/v_0^2}\right) && v<v_{\rm esc}, \\
    &0 && v\geq v_{\rm esc},
  \end{aligned} \right.
\label{eq:SHM}
\end{equation} 

\noindent
where $N$ is a normalization factor. The local circular speed is usually assumed to be $v_0=220$~km/s and the commonly adopted escape speed is $v_{\rm esc}=544$~km/s.

A departure from this commonly assumed model changes the predicted event rate in direct detection experiments. In fact, the true dark matter distribution could be different from what is normally assumed. One way to access the information on the properties of the DM halo is to examine the dark matter distribution of the MW analogues in the cosmological simulations. For example, the DM-only simulations of structure formation have shown that the velocity distributions have substantial deviations from a Maxwellian shape \cite{Green:2000jg,Green:2002ht,Vogelsberger:2008qb,Kuhlen:2009vh}. Nevertheless, in recent years by using more realistic high resolution cosmological simulations, which include baryonic physics effects, it was shown that the dark matter velocity distributions actually agree rather well with the SHM. Yet, the local circular speed of the Maxwellian distribution and the escape velocity may well be different \cite{Bozorgnia:2017brl,Bozorgnia:2016ogo,Kelso:2016qqj,Sloane:2016kyi,Bozorgnia:2019mjk}. In this section, we focus on investigation of the MW-like galaxies from the IllustrisTNG project. We first introduce the general properties of the used simulation and describe the selection criteria used to identify the MW analogues. We then present the resulting velocity distribution functions for some selected halos.

\subsection{Selection of MW-like galaxies}
\label{sec:illustris_selection}

The IllustrisTNG project is a publicly available suite of state-of-the-art magneto-hydrodynamic cosmological simulations \cite{Pillepich:2017jle,Nelson:2017cxy,Springel:2017tpz}, an extension to the Illustris simulations \cite{Genel:2014lma,Vogelsberger:2014dza,Sijacki:2014yfa}. The cosmological parameters were chosen to be those derived from the analysis of the Plank Collaboration~XIII \cite{Ade:2015xua}, namely $\Omega_{\rm{m}}~=~0.3089,\,\Omega_{\rm{b}}~=~0.0486,\,h~=~0.677,\,\sigma_8~=~0.8159$ and $n_{\rm{S}}~=~0.9667$. IllustrisTNG is evolved using the moving mesh hydrodynamics code AREPO \cite{Springel:2017tpz} and it contains a number of improvements of sub-grid physics models (e.g. models for stellar and AGN feedback, as well as black hole growth \cite{Pillepich:2017fcc}) with respect to the original Illustris project. These simulations are able to reproduce many key relations in the observed galaxies, such as the stellar mass function and the size distribution at low and high redshift, the fraction of dark matter within galaxies at $z=0$. For more details, see \cite{Nelson:2017cxy,2019MNRAS.484.5587T,2018MNRAS.474.3976G,Lovell:2018amb,Vogelsberger:2017sjl,10.1093/mnras/sty2078}. Furthermore, as it was shown in \cite{Lovell:2018amb}, the MW-mass galaxies of the IllustrisTNG reproduce adequately the observed MW rotation curve, this suggests that the matter distribution in the simulated MW-mass galaxies is realistic and the baryonic effects in the dark matter are properly modelled. That makes them of particular relevance for our study. 

In this paper we make use of the highest (to date) resolution version of the IllustrisTNG project, which is TNG100. From the suite of TNG100 simulations we select candidate MW analogues in a $(75/h\approx110.7\,\rm{Mpc})^3$ box. The baryonic and dark matter mass resolutions are $m_{\rm baryon}=1.4\times10^6\,M_{\odot}$ and $m_{\rm DM}=7.5\times10^6\,M_{\odot}$, respectively. 

In order to identify simulated galaxies which satisfy MW observational constraints we adopt the following criteria:

 \begin{itemize}
     \item we identify all galaxies in the mass range\footnote{These halo mass cuts were selected to better match MW-like galaxies in TNG100 simulations, as stressed in \cite{2019MNRAS.486.4686K}.} $5\times10^{11}\leqslant M_{200}/M_{\odot}\leqslant 2\times10^{12}$, where $M_{200}$ is defined as the mass enclosed within the sphere that contains a mean density 200 times the critical density;
     \item within 2456 galaxies identified in the previous step we select the ones that have the stellar mass range $\rm  2.5\times10^{10}\leqslant M_{*}/M_{\odot}\leqslant 5\times10^{10}$ and the gaseous mass to stellar mass fraction range $\rm  0.03\leqslant M_{gas}/M_{*}\leqslant 1$. At this step we end up with 340 galaxies;
     \item among those 340 galaxies we identify galaxies that have a substantial stellar disc component. In order to do so we select only galaxies that contain more than 50\% of stars with the circularity parameter $\epsilon>0.7$, i.e., those stars that have  significant (positive) rotational support. Here we also require that selected galaxies do not have any significant merger after redshift $z=0.68$ and we end up with 164 galaxies.

 \end{itemize}

For each galaxy in the sample, we align the coordinate system with the principal axes. We then select DM particles in torus, along the plane in the stellar disc, around the solar radius, $ R_{\odot}=8\,\rm{kpc}$, with both radial and vertical extents $\pm$2~kpc. This region contains a total of $\sim$~700-900 dark matter "particles".

\subsection{Dark matter velocity distributions}
\label{sec:illustris_velocity_functions}

\begin{figure}
    \centering
    \includegraphics[width=0.95\textwidth]{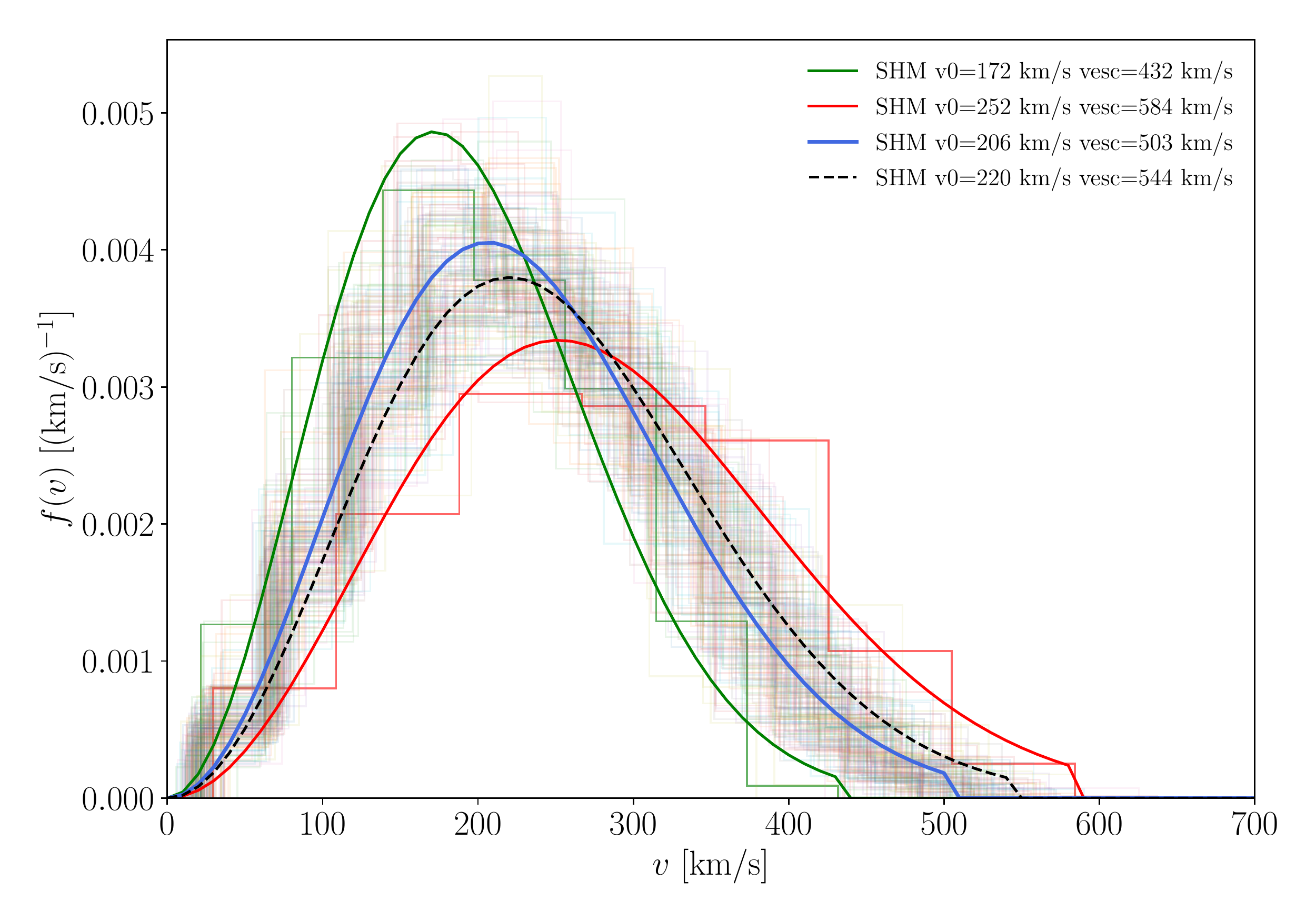}
    \caption{The histograms of the local dark matter speed distribution in the Galactic rest frame for the 164 hydrodynamical TNG100 Milky-Way-like halos selected by our analysis.
    The red, blue and green solid lines represent the best-fit Maxwellian distribution for the minimum (MIN), median (MED) and maximum (MAX) peak speed values, respectively. The histograms corresponding to the MIN and MAX distributions are highlighted.}
    \label{fig:simul}
\end{figure}

To derive the dark matter distribution we compute the average speed distribution of DM particles in the  torus defined above. The speed distribution, $f(v)$, is normalised to unity as $\int dv f(v)=1$, and is related to the velocity distribution, $\tilde{f}({\bf v})$, by
\begin{equation}
    f(v) = v^2\int d\Omega_{\bf v}\tilde{f}({\bf v}),
    \label{eq:vel_funciton}
\end{equation}
\noindent
such that $\int d^3v\tilde{f}({\bf v})$ = 1. Here $d\Omega_{\bf v}$ is an infinitesimal solid angle around the direction $\bf v$.

In Fig.~\ref{fig:simul} we present the local dark matter speed distributions in the Galactic rest frame for the MW-like halos of TNG100 simulations. As can be seen in the figure, there is quite large halo-to-halo  variation in the speed distribution. The SHM peak speeds are in the range of $v_0$=[172-252]~km/s and the escape speeds are in the range $v_{\rm esc}=[379-637]$~km/s. The peak speeds of the distribution are found by fitting Eq.~(\ref{eq:SHM}) while the escape speed is taken to be the highest speed of any particle in the entire halo. 

The derived ranges are consistent with, though somewhat larger than, those found, for example, in \cite{Bozorgnia:2019mjk}. The relatively large spread in the speed distributions owes to the fact that in this work we find many more halos that fulfill the selection criteria due to using a lower resolution simulation.
 Nevertheless, it would be interesting to cross-check our results within the same suite of simulations but using a higher resolution box. This can become possible once the results of TNG50 become publicly available \cite{Pillepich:2019bmb}. 

In order to study the impact of inferred ranges of the speed distributions on the direct DM searches we choose the halos with the minimum and the maximum peak speeds, shown in Fig.~\ref{fig:simul} in green and red, respectively. We also define the median values of the peak speed and the escape velocity, the corresponding velocity distribution is shown in Fig.~\ref{fig:simul} in blue. We refer to them as MIN, MED and MAX profiles, as they correspond to minimal, median and maximal expected strength of the signal in direct detection experiments. We present the results in Table~\ref{tab:simul_profile}. Additionally, for each case we calculate the local dark matter density in the torus region, that we also list in Table~\ref{tab:simul_profile}. The resulting local velocity profiles are first of our main results of this work. 

\begin{table}[h] 
\centering
\addtolength{\tabcolsep}{6pt}
\scalebox{1}{%
\begin{tabular}{ | c | c | c | c |}
\hline 
Name & $v_0$ [km/s] & $v_{\rm esc}$ [km/s] & $\rho_{\chi}\,{\rm [GeV/cm^3]}$\\
\hline 
MIN & 172 & 432 & 0.27 \\[0.6ex] 
MED & 206 & 503 & 0.41 \\[0.6ex] 
MAX & 252 & 584 & 0.32 \\[0.6ex] 
\hline
\end{tabular}}
\caption{The best-fit values of the Maxwellian distribution of the local DM speed distribution and the corresponding escape velocity as well as average dark matter density in the torus region of the TNG100 MW-like halos.}
\label{tab:simul_profile} 
\end{table}

\section{Uncertainties in sub-GeV DM electron recoil}
\label{sec:subGeV}

As an illustrative and timely example we apply the DM speed distribution profiles found in the previous section to direct detection searches of sub-GeV dark matter. The impact of astrophysical uncertainties on the sub-GeV dark matter range is a largely unexplored area, with most studies so far focusing on WIMP detection through standard nuclear recoil experiments \cite{McCabe_2010,Frandsen_2012,FRIEDLAND2013183,Benito:2016kyp,Green_2017,Ibarra:2018yxq,Belyaev:2018pqr}. These studies accommodated a wide range of circular and escape velocities compatible with the calculations given in the Illustris simulation in this work.\footnote{Note, however, that the escape velocity from our IllustrisTNG-inferred MIN profile is substantially lower which can have some non-negligible impact on limits on low-mass WIMPs in nuclear recoil based experiments as well.}
However, nuclear recoil experiments are insensitive to dark matter in the sub-GeV range. For such light DM the most promising detection techniques are based on electron recoils, which can provide a strong signal through electron ionization in small band-gap semiconductors. For silicon and germanium this gap is typically $\sim$1 eV, and so allows for detection of dark matter even as light as hundreds of keV.

The scattering DM-electron rate in a semiconductor, assuming that DM speed distribution  $f(v)$ follows a symmetric MB distribution, is given in events/kg/day by \cite{Essig:2015cda}
\begin{eqnarray}
    R_{\text {crystal}}=\frac{\rho_{\chi}}{m_{\chi}} N_{\text {cell }} \bar{\sigma}_{e} \alpha  \frac{(m_{e}+m_\chi)^{2}}{m_{\chi}^{2}} \!\int\!\! d \ln E_e d \ln q\left(\frac{E_{e}}{q} \eta\left(v_{\min }\left(q, E_{e}\right)\right)\right) F(q)^{2}\left|f_{\text 
    {crystal}}\left(q, E_{e}\right)\right|^{2},
\end{eqnarray}
where $\rho_{\chi}$ and $m_{\chi}$ are the local density and mass of the DM particle, respectively, and $m_e$ is the mass of the electron. The crystal form factor, $f_{\text{crystal}}\left(q, E_{e}\right)$ contains information about the electronic structure of the material, $N_{\text {cell }}$ is the number of unit cells in the crystal target, $\alpha$ is the fine structure constant and $\bar\sigma_e$ parameterizes the strength of the interaction, which in the case of $F(q)=1$ is equal to the cross-section for an elastic scattering of DM on a free electron.

To calculate the rate for the detector we must first boost the DM velocity with the average Earth velocity relative to the DM halo, taken to be $v_E = 240$ km/s such that we define $g(\textbf{v}) \equiv f(\textbf{v}+\textbf{v}_E)$. With this consideration, the astrophysical properties of the halo are then encapsulated by the function $\eta\left(v_{\min }\right)$ such that
\begin{equation}
    \eta\left(v_{\min }\right)=\int \frac{d^{3} v}{v} g(v) \Theta\left(v-v_{\min }\right),
\end{equation}

where, through energy conservation, $v_{\min}$ is a function of the electron energy $E_e$ and momentum transfer $q$,
\begin{equation}
\label{eq:vmin}
    v_{\min }\left(q, E_{e}\right)=\frac{E_{e}}{q}+\frac{q}{2 m_{\chi}}.
\end{equation}
This determines the minimum speed DM particles that is required to excite electrons in the crystal, transferring $q$ momentum and energy $E_e$. The stark contrast between electron excitation and nuclear recoil based experiments consists in the fact that the electron is a fast and light particle. Therefore,
the momentum transfer $q$ will be approximately given by the momentum of the electron $q\approx m_e v_e$, where $v_e$ is the electron velocity which is typically an order of magnitude larger than the circular velocity of the halo, $v_0$. Hence, energy transitions to states above the band gap are sensitive to the distribution around the tail of the DM velocity profile. 

The interaction between an incoming DM particle and an electron in the crystal is in fact a rather complicated process. It is understood that after the DM particle deposits the initial energy $E_e$ to the electron, additional secondary scatterings redistribute this energy between lower-lying electron or hole energy states. Eventually the secondary scatterings fall below a certain energy threshold for electron-hole pair production, $\varepsilon$, and the number of total pairs generated is what is measured. Assuming pair creation is linear in energy, and following the conventions in \cite{Essig:2015cda}, we define the \textit{ionization threshold} $Q$  as a function of the electron energy $E_e$,
\begin{equation}
    Q = 1+\left\lfloor\frac{E_e - E_{\text{gap}}}{\varepsilon}\right\rfloor ,
    \label{eqn:Q}
\end{equation}
where the energy band gaps $E_{\text{gap}}$ for silicon and germanium are 1.11 eV and 0.67 eV, respectively. The energy needed to produce an additional electron-hole pair above the band gap, denoted by $\varepsilon$, is also determined to be 3.6 eV and 2.9 eV for silicon and germanium, respectively. The expression (\ref{eqn:Q}) takes only integer values for $Q$ since it represents the total number of electron-hole pairs produced for a single DM interaction. This is a key parameter in calculating experimental sensitivity curves for direct detection.

For example, the DAMIC experiment that is currently running at SNOLAB \cite{Chavarria:2014ika} expects to reach a threshold as low as 2-3 electrons in a total of 1 kg of detector with its current level of leakage current (produced by thermal excitation in the crystal) demonstrated to be very low in the prototype run \cite{Settimo:2018qcm}. Among other considerations, improvements in leakage current reduction through operation temperature and material quality is imperative in reaching low thresholds. Hence, in this study we examine the consequences of halo DM properties on multiple thresholds.

To calculate the direct detection limits for silicon and germanium targets, we employ the publicly available code \texttt{QEdark} \cite{Essig:2015cda} in accordance with the plane wave self-consistent field (\texttt{PWscf}) code in \texttt{Quantum ESPRESSO} \cite{Giannozzi_2009} for the electron wave-functions and energy levels. \texttt{QEdark} is utilized to numerically compute the crystal form factors for both cases, and then subsequently the total scattering rate and sensitivity curves.

\begin{figure}
    \centering
    \includegraphics[width=0.475\textwidth]{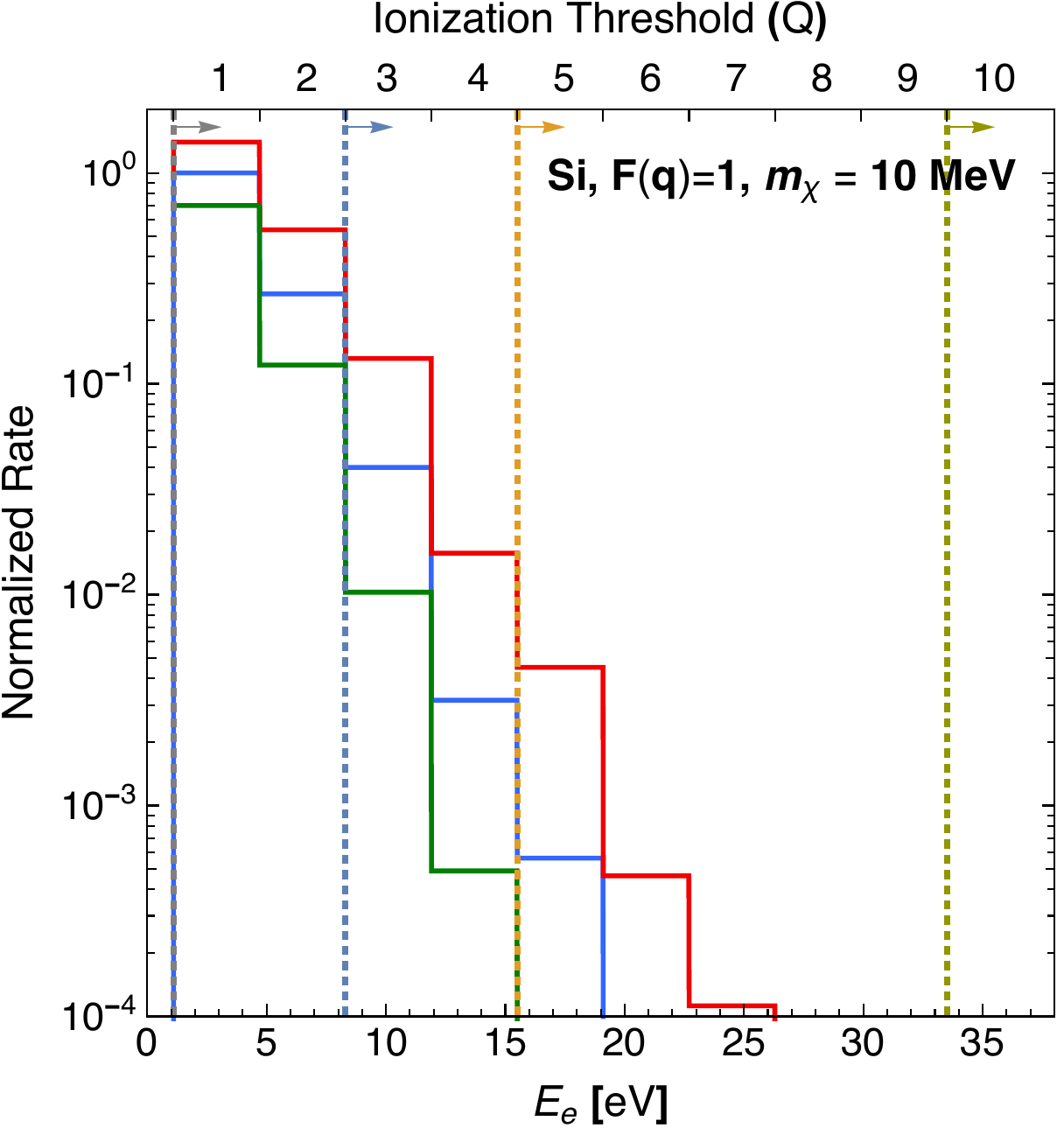}
    \includegraphics[width=0.475\textwidth]{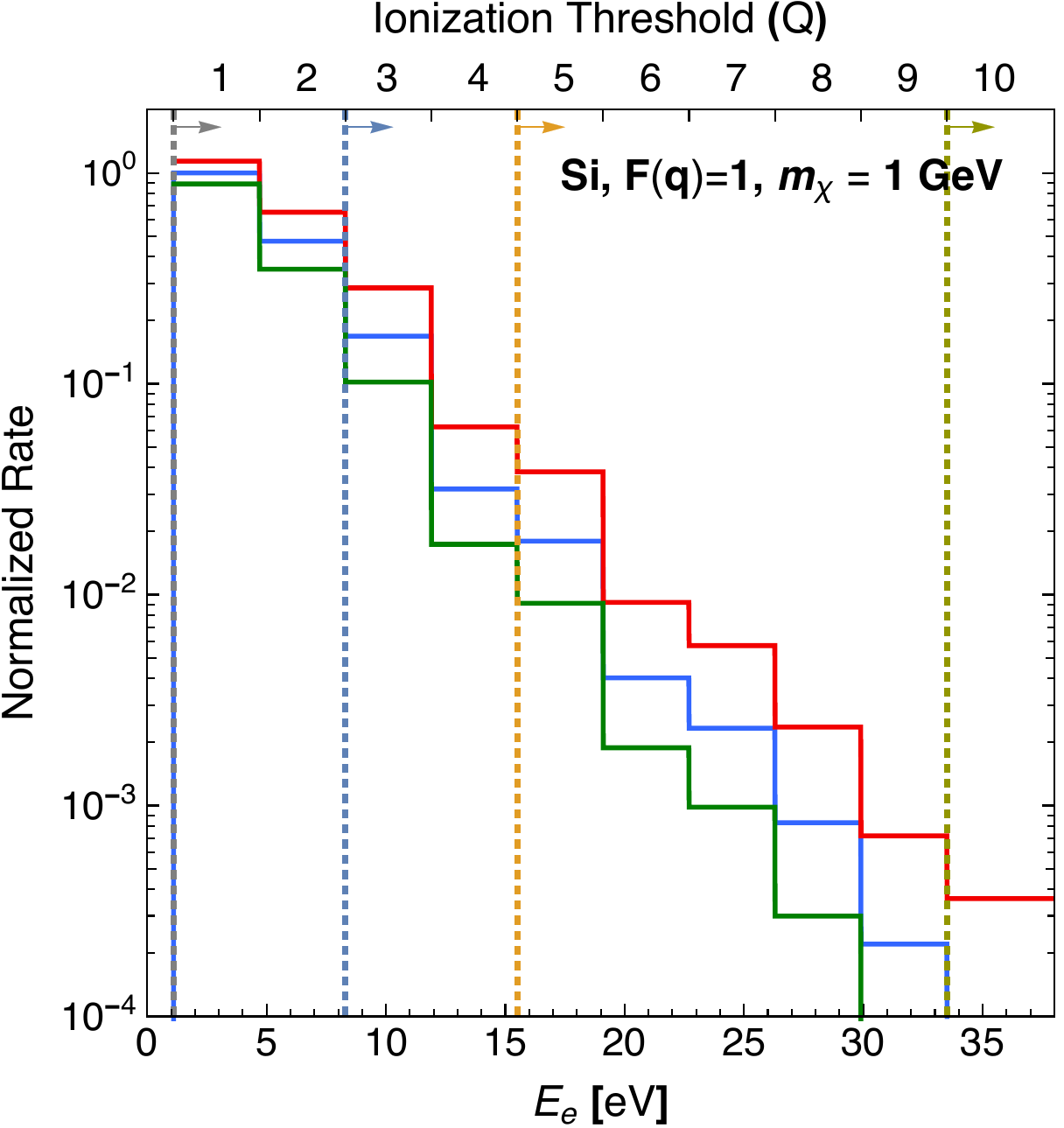}
    
    \vspace{0.5cm}
    \includegraphics[width=0.475\textwidth]{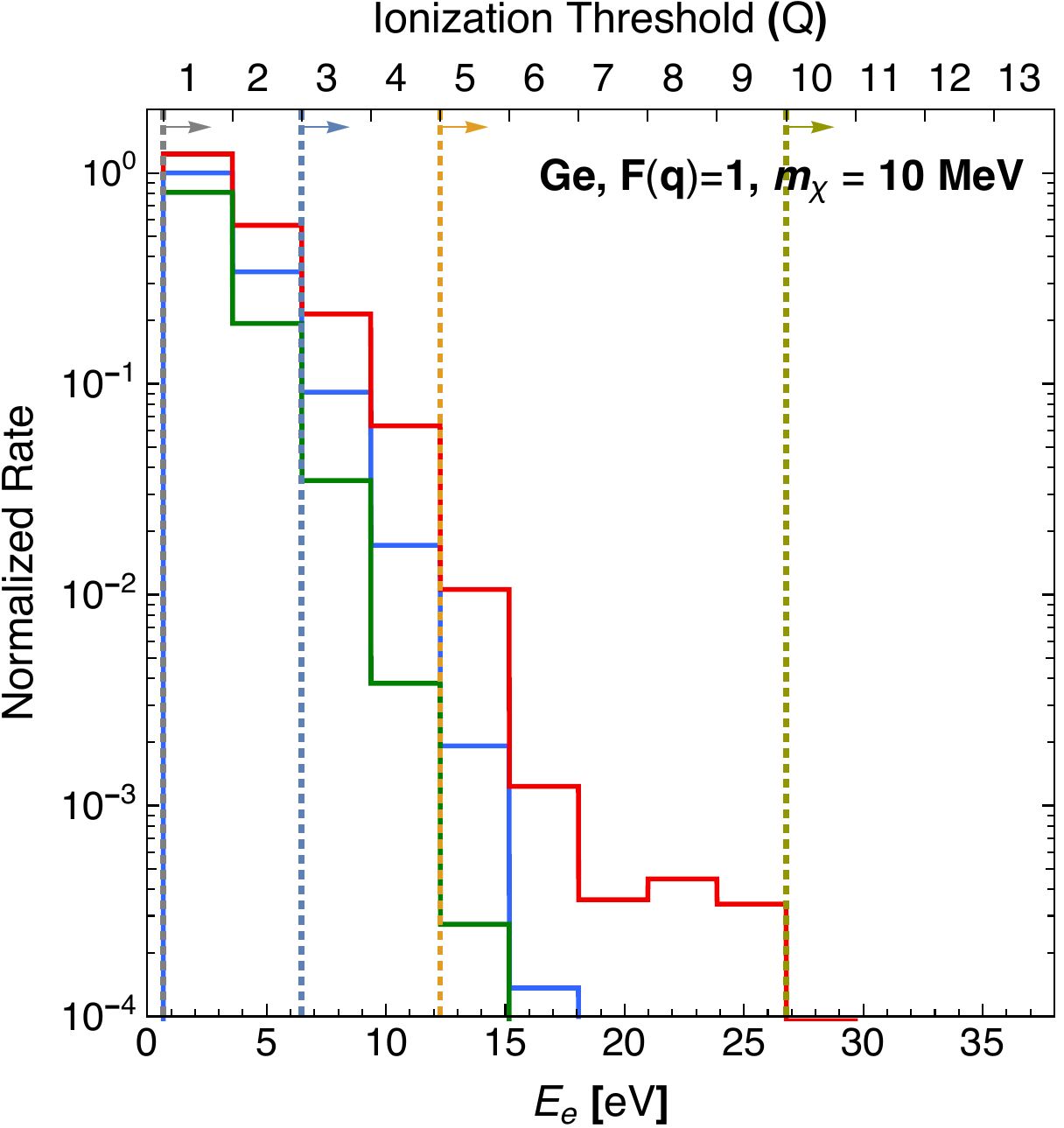}
    \includegraphics[width=0.475\textwidth]{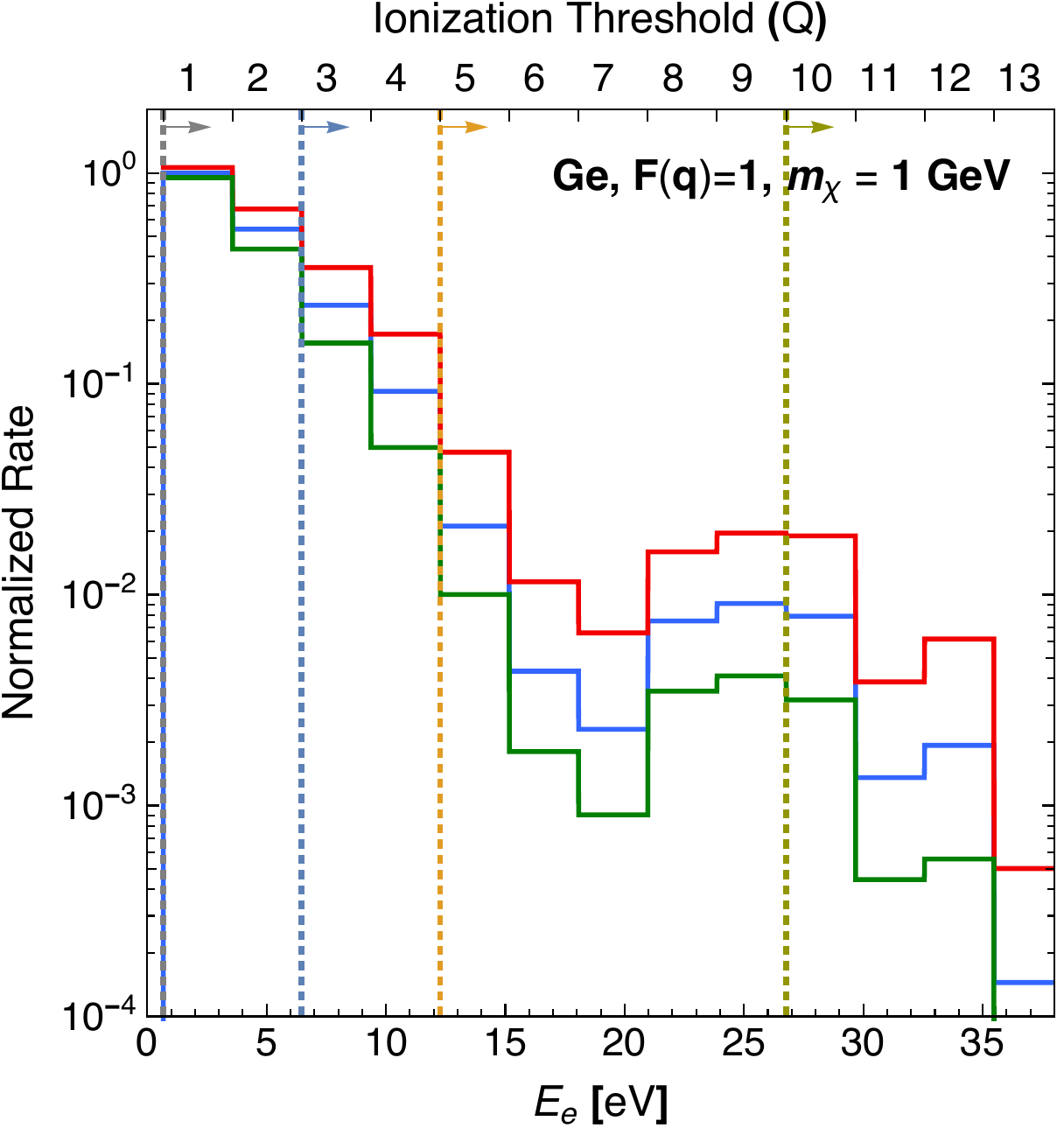}
    \caption{Normalized event rate vs. electron energy and ionization threshold (number of electron-hole pairs). The top panels show germanium target, while the bottom ones silicon target, whereas the left ones are for $m_\chi$=10~MeV and the right ones for $m_\chi$=1~GeV. The color coding is as in Fig.~\ref{fig:simul}, corresponding to the central MED (blue), MAX (red) and MIN (green) velocity profiles. All events are normalized to the first bin of the MED distribution. The vertical lines indicate the electron-hole pair ionization thresholds as will be shown below in Figs.~\ref{fig:Si1} and \ref{fig:Ge1}; only electron energies to the right of a given line are sufficient to produce a given number of pairs. }
    \label{fig:rates}
\end{figure}

To illustrate the impact of different velocity profiles in Fig.~\ref{fig:rates} we compare normalized event rates for the MIN (green), MED (blue) and MAX (red) cases, for Ge and Si targets and two different DM mass values. The large difference in the  velocity tails of these profiles, parametrized mostly by the value of the $v_0$, translates to more extended tails in the electron energy distribution $E_e$. This leads to higher expected rates, with the difference being more pronounced in the higher ionization thresholds. The characteristic shape of the rate histogram for germanium (bottom right panel) is a result of the fast-moving 3d-shell electrons that dominate at around $E_e \gtrsim 24$~eV. Despite interacting with a larger momentum transfer with the DM particle, their contributions are not significant for the smaller thresholds.

\begin{figure}
    \centering
    \includegraphics[width=0.475\textwidth]{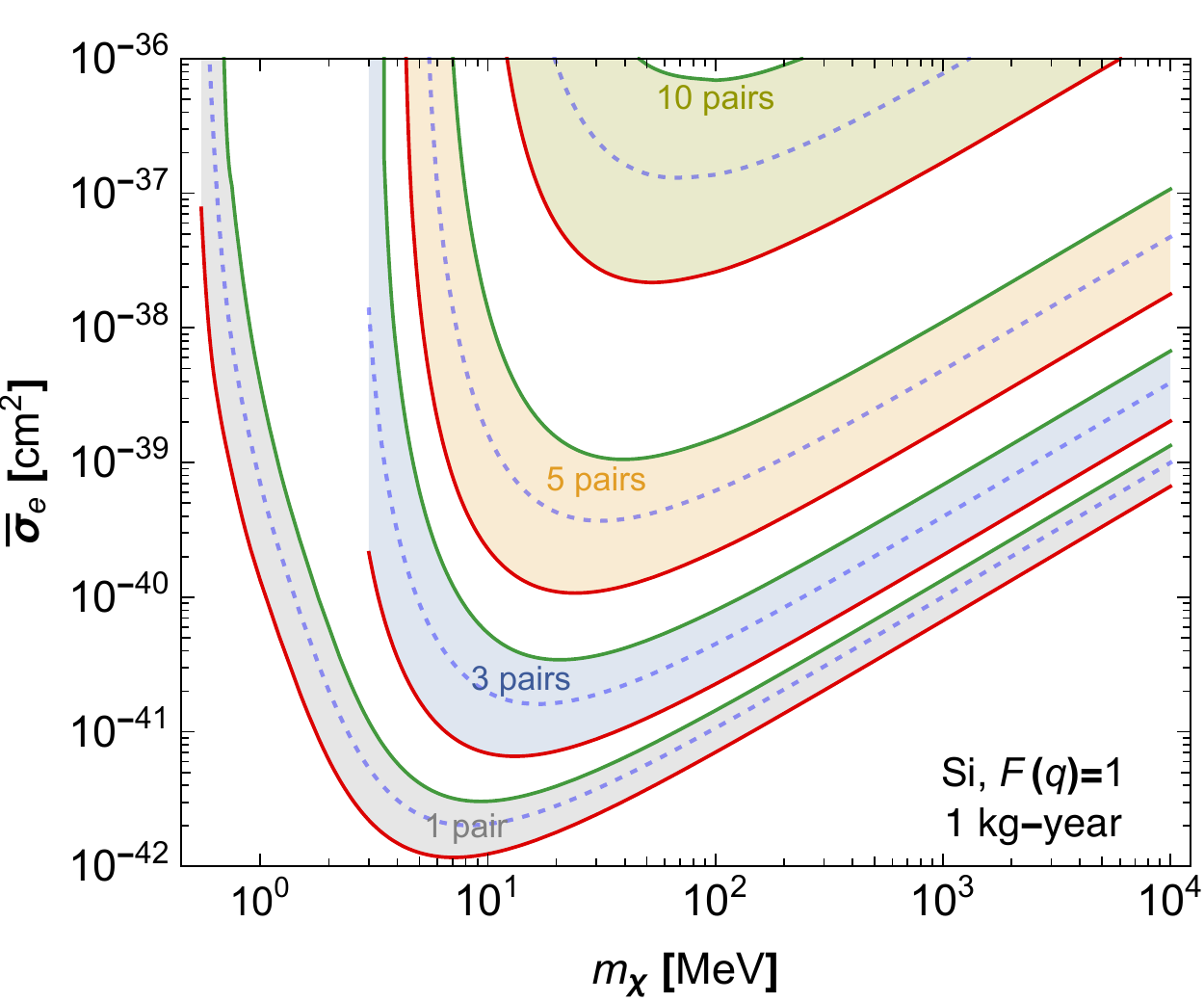}
    \hspace{0.1cm}
    \includegraphics[width=0.495\textwidth]{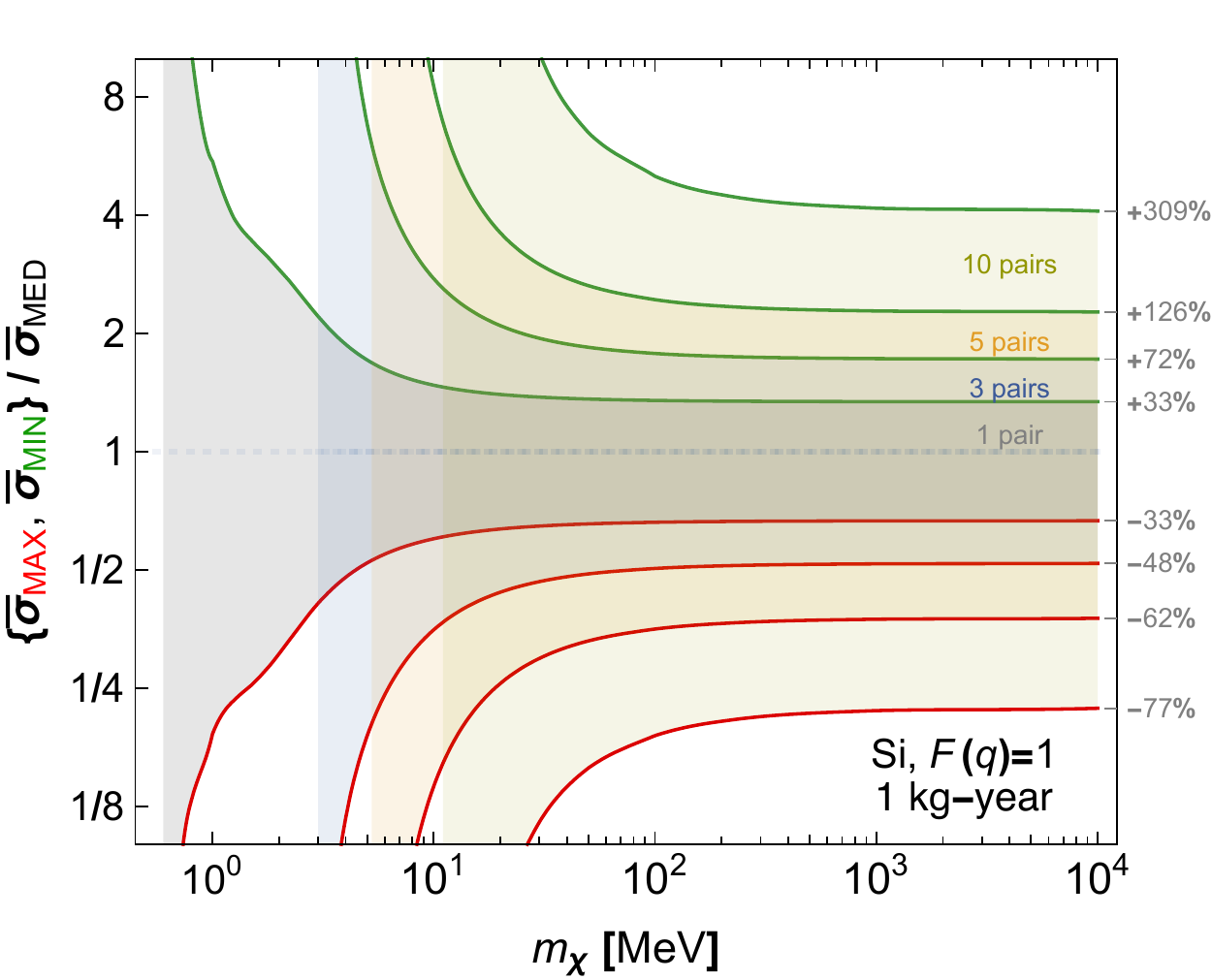}
    \caption{The projected 95\% CL sensitivities (left panel) and the size of the astrophysical uncertainty (right panel) for a  silicon target with DM interaction described by the form factor $F(q)=1$. The results are given for four different values of ionization threshold $Q$: 1 (gray), 3 (blue), 5 (orange) and 10 (light green) electron-hole pairs. In all cases the limiting solid lines correspond to the MIN (green) and MAX (red) velocity profiles, while the dotted lines show the central MED profile. In all the cases the difference in the local DM density is taken into account. In the right panel the size of the uncertainty is shown as the ratio of the MIN and MAX profiles to the MED profile, and on the right hand side of the frame also as the percentage difference from the best-fit case.
    }
    \label{fig:Si1}
\end{figure}

\begin{figure}
    \centering
    \includegraphics[width=0.475\textwidth]{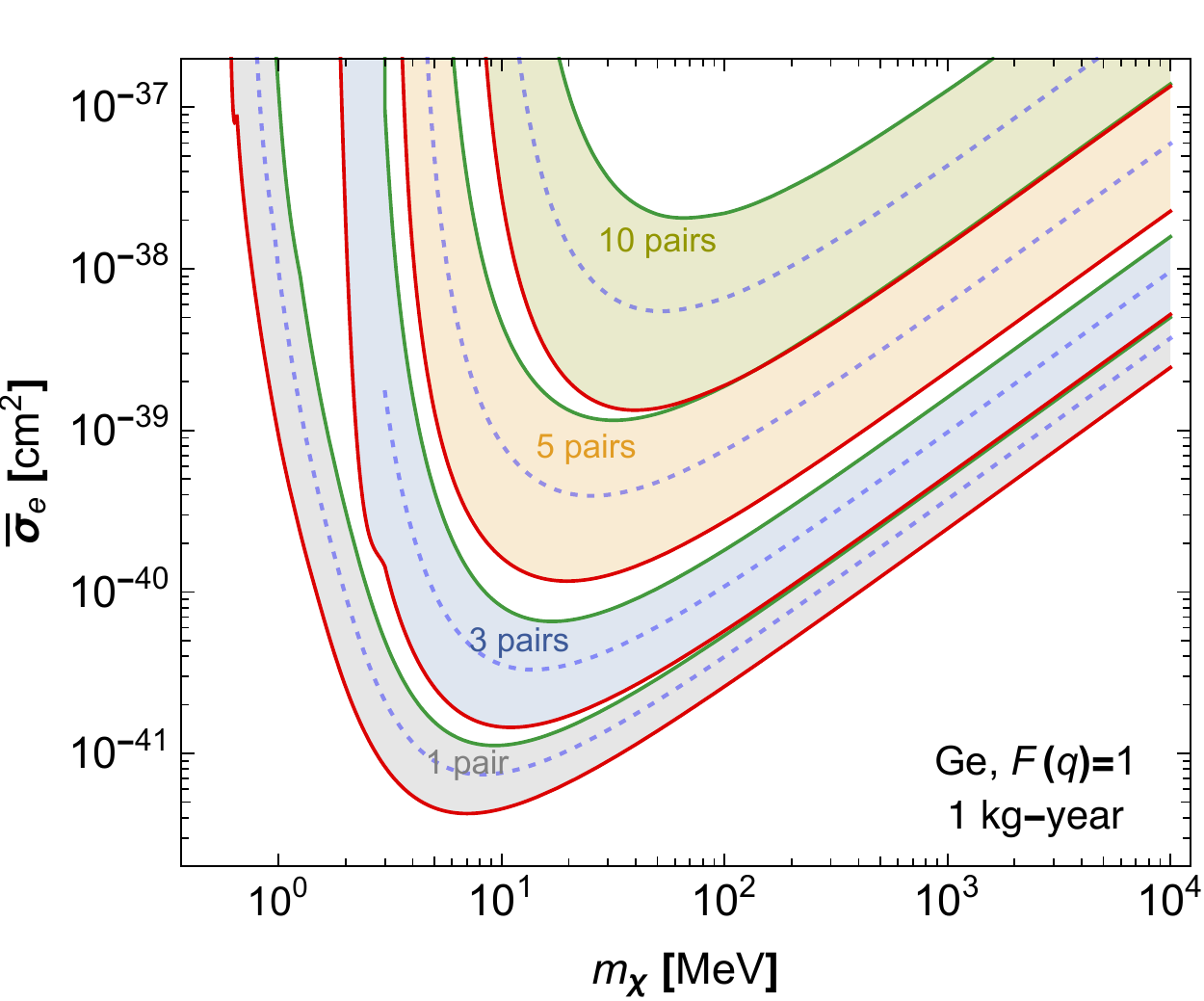}
    \hspace{0.1cm}    \includegraphics[width=0.495\textwidth]{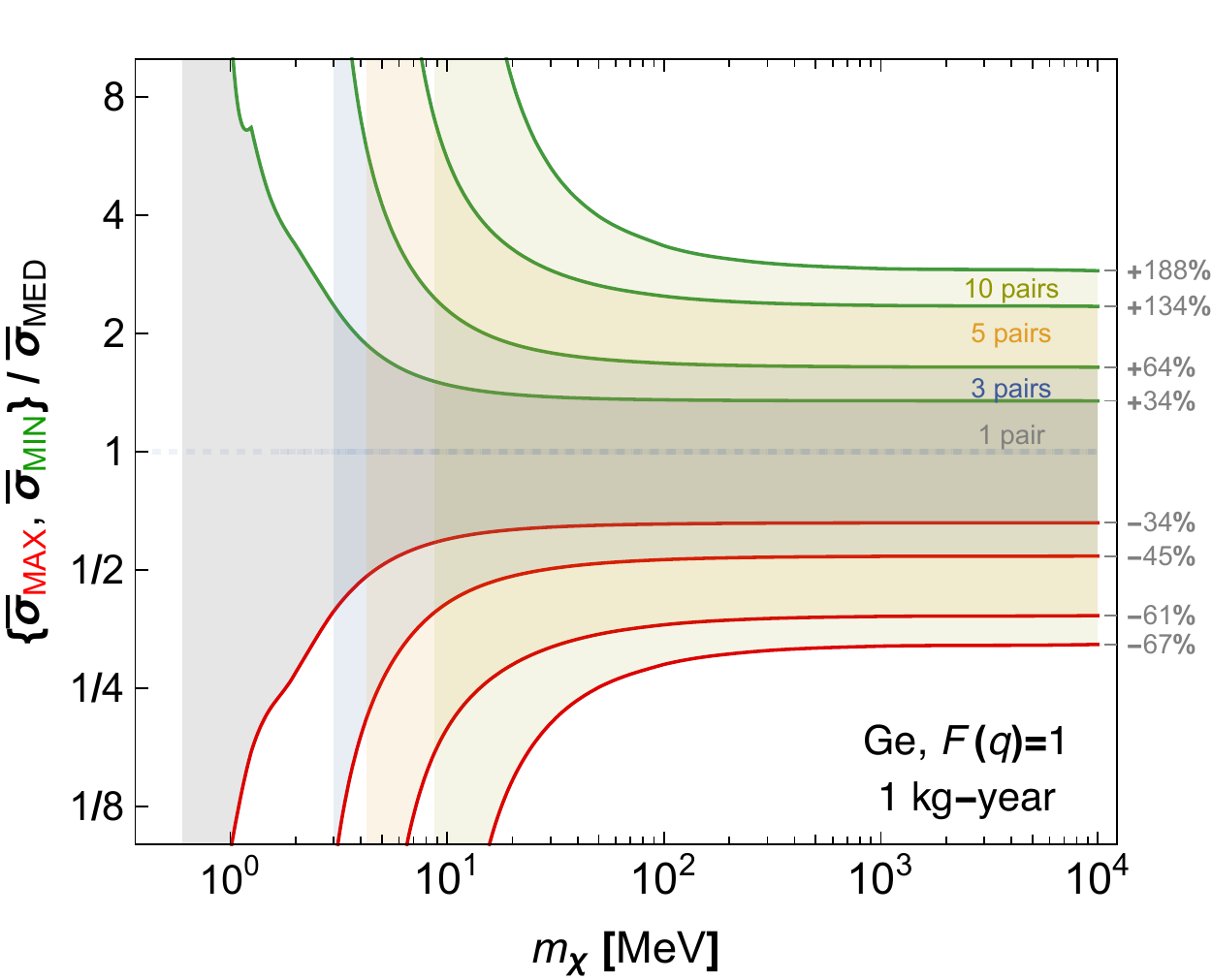}
    \caption{The same as Fig.~\ref{fig:Si1} but for germanium target.}
    \label{fig:Ge1}
\end{figure}

The final results for projected 95\% CL sensitivities for an 1 kg-year exposure with the astrophysical uncertainties taken into account are given in Fig.~\ref{fig:Si1} for silicon and in Fig.~\ref{fig:Ge1} for germanium targets. Left panels show the projected limits on the cross-section $\bar\sigma_e$ and right show the relative size of the uncertainty due to the DM density and velocity profile. In the limit of very large $m_\chi$ the uncertainty reaches a constant, since the minimal velocity essentially becomes independent of the DM mass, see eq. (\ref{eq:vmin}). For light DM masses near the detection thresholds the uncertainty in the sensitivity grows significantly where both the circular and escape velocity of the DM profile contribute sizable effects. This shifts the precise position of the minimum reachable mass for a particular pair threshold. The projected sensitivities are calculated under the assumption of no background, with the understanding that it is effectively parameterized by the value of $Q$ that is achievable (for more detailed discussion of the backgrounds see \cite{Essig:2015cda}).

What is worth stressing is that: (a) the impact of the velocity profile is quite significant, and is comparable to the change between consecutive electron-hole pair thresholds, and (b) the lower the electron-hole pair threshold is experimentally achievable, the better is not only the overall sensitivity but also the robustness of the detection method.

\section{Discussion and conclusions}
\label{sec:conclusions}

In this paper we analyzed the results of the recent IllustrisTNG hydrodynamical simulation and obtained the local DM halo properties. We identified the resulting best fit (MED) as well as -- from the point of view of predicted sensitivity of DM direct detection experiments -- extreme MIN and MAX profiles. These profiles amount to first of the two main results of this work and can be adopted in updated direct detection analyses. Next we used them to quantify, for the first time, the astrophysical uncertainties of the sub-GeV electron recoil DM searches.

Qualitatively, based on experience with nuclear recoil based searches, one would expect that: (a)~the impact of the velocity profile is to shift the DM minimal mass threshold, and (b) that the effect is most pronounced for $m_\chi \sim m_e$. In many electron recoil based detection techniques, however, because of their sensitivity to the energy deposition in generating the measurable signal even for $m_\chi \gg m_e$, the velocity profile of DM can play a significant role. We expect this to be the case for not only semiconductor-based detectors, but potentially also detectors using Dirac materials or superconductors. In contrast, experiments measuring only the direct effect of single recoil or absorption, without sensitivity to the exact value of the energy transfer, suffer from much smaller astrophysical uncertainties, more closely resembling the known case of nuclear recoils.

Quantitatively, our results show that the uncertainty reaches more than an order of magnitude for two most studied semiconductor targets, silicon and germanium. More precisely, even in the limit of large DM mass ($m_\chi \gtrsim 100$ MeV), the sensitivity difference between MIN and MAX velocity profiles reaches factor $\mathcal{O}(10-20)$ depending on the target material and more importantly the ionization threshold. This is even for the contact-like interaction form factor $F(q)=1$, while other form factors, favoring interactions with smaller momentum transfer, $F(q)\sim 1/q$ or $F(q)\sim 1/q^2$, lead to larger effect of $\mathcal{O}(15-35)$. Moreover, for lower DM masses the effect is much larger, due to higher sensitivity of the tail of the velocity distribution.

Therefore, to summarize, the second main result of this work underlines how crucial it is to develop background reduction techniques for experiments based on electron recoils, as not only the sensitivity grows and theoretical uncertainty diminishes, but also the impact of the not yet well known dark matter halo profile is much less severe if one is able to be sensitive to the lowest possible ionization thresholds.


\acknowledgments

E.~K. thanks Ewa {\L}okas and Ivana Ebrova for a number of insightful discussions on the matter of the Illustris simulations. A.~H. is supported in part by the National Science Centre, Poland, research grant No. 2018/31/D/ST2/00813. E.~K., L.~R. and M.~T. are supported by the grant ``AstroCeNT: Particle Astrophysics Science and Technology Centre" carried out within the International Research Agendas programme of the Foundation for Polish Science financed by the European Union under the European Regional Development Fund. L.~R. is also supported in part by the National Science Centre research grant No. 2015/18/A/ST2/00748. 

\bibliographystyle{JHEP.bst}
\bibliography{biblio.bib}

\end{document}